\documentclass[12pt,preprint]{aastex}

\newcommand{\zem}{\mbox{$z_{\rm em}$}}

\begin{document}


\title{Variability-selected quasars behind the Small Magellanic
Cloud\altaffilmark{1}}

\author{A. Dobrzycki\altaffilmark{2}, L. M. Macri\altaffilmark{3,4},
K. Z. Stanek\altaffilmark{2}, and P. J. Groot\altaffilmark{5}}

\altaffiltext{1}{Based on observations collected at the Magellan Baade
6.5-m telescope.}

\altaffiltext{2}{Harvard-Smithsonian Center for Astrophysics, 60
Garden Street, Cambridge MA 02138, USA,
[adobrzycki,kstanek]@cfa.harvard.edu.}

\altaffiltext{3}{Kitt Peak National Observatory, National Optical
Astronomy Observatory, 950 North Cherry Avenue, P.O. Box 26732,
Tucson, AZ 85726-6732, USA, lmacri@noao.edu.}

\altaffiltext{4}{Hubble Fellow}

\altaffiltext{5}{Department of Astrophysics, University of Nijmegen,
PO Box 9010, 6500 GL Nijmegen, The Netherlands, pgroot@astro.kun.nl.}

\shorttitle{Variability-selected quasars behind the SMC}
\shortauthors{Dobrzycki et al.}

\slugcomment{To appear in the Astronomical Journal, March 2003 issue}


\begin{abstract}

We present followup spectroscopic observations of quasar candidates in
the Small Magellanic Cloud selected by Eyer from the OGLE database. Of
twelve observed objects identified as ``QSO Candidate'', five are
confirmed quasars, with the emission redshifts ranging from 0.28 to
2.16. Two of those quasars were also recently identified independently
in the MACHO database by Geha et al. We discuss the prospects of using
variability-based selection technique for quasar searches behind other
dense stellar fields. An additional criterion utilizing the
color-color diagram should reduce the number of stars in the candidate
lists.

\end{abstract}

\keywords{Magellanic Clouds --- quasars: individual}


\section{Introduction}

Searches for quasars in dense stellar fields --- such as the
Magellanic Clouds --- were in the past hampered by difficulties in
selecting candidates. Optical followup on X-ray selected objects had
to deal with large number of candidates in X-ray source error boxes,
while variability studies required monitoring of vast number of
objects. At the same time, such quasars are of great astrophysical
interest, for example as reference points for analysis of proper
motions and as background sources for absorption studies. Until
recently, only a handful of confirmed quasars behind the LMC and SMC
were known (Blanco \& Heathcote 1986; Crampton et al.\ 1997; Tinney et
al.\ 1997; Anguita, Loyola, \& Pedreros 2000; see also Kahabka et al.\
1999, Haberl et al.\ 2000 and Kahabka, de Boer, \& Br\"uns
2001). Almost all of those quasars were located behind the outer,
sparse parts of the Clouds.

Recent developments --- such as the launch of the {\em Chandra X-ray
Observatory\/} with its superb spatial resolution in X-rays, and
availability of large photometry databases (OGLE, MACHO) --- are now
making systematic searches for quasars in dense stellar fields
possible. Recently, Dobrzycki et al.\ (2002) found four X-ray quasars
among serendipitous sources in four {\em Chandra\/} observations of
objects in the LMC coinciding with the OGLE fields.

A characteristic that was often used in quasar surveys was their
irregular variability. Several such surveys were performed or are
on-going (Hawkins 1983; Meusinger \& Brunzendorf 2001; Rengstorf et
al.\ 2001). In the Magellanic Clouds, this technique has been used by
Geha et al.\ (2002), who published a list of forty seven quasars
behind the LMC and SMC, selected from the MACHO database.

Between 1997 and 2001, large parts of the Magellanic Clouds were
monitored for microlensing events by the Optical Gravitational Lensing
Experiment (OGLE-II: Udalski, Kubiak \& Szyma\'nski 1997). Udalski et
al.\ (1998) released $BVI$ photometry and astrometry of 2.2 million
objects from the central parts of the SMC\footnote{Data available from
ftp://bulge.princeton.edu/ogle/ogle2/maps/smc/}. In addition, a large
catalog of 68,000 variable objects observed by OGLE-II in both the LMC
and the SMC was prepared by \.Zebru\'n et al.\ (2001)\footnote{Data
available from http://bulge.princeton.edu/$\sim$ogle/ogle2/dia/},
based on a version of the image subtraction software (Alard \& Lupton
1998) developed by Wo\'zniak (2000).

Eyer (2002) presented an algorithm for selecting quasar candidates
from objects in the OGLE database. He searched the database for slowly
and irregularly varying blue objects and identified QSO candidates
(``QCs'') towards the Magellanic Clouds: 118 QCs towards the LMC and
15 towards the SMC. Eyer also identified several ``Unclassified''
objects, which had similar light curve characteristics.

In 2002 September, we performed followup observations of twelve
brightest ``QSO Candidates'' from the SMC from the Eyer's list with
the Magellan 6.5-meter Baade telescope. Five of them turned out to be
previously unknown quasars; an excellent success rate. We also
observed all four of Eyer's ``Unclassified'' objects in the direction
of the SMC and none of them turned out to be a QSO.

Coincidentally, less then a week after our observations had been
completed, the paper by Geha et al.\ (2002) was posted on astro-ph. In
this paper, the authors performed an analysis of the MACHO variable
star database. They selected 360 quasar candidates behind the
Magellanic Clouds, and completed followup observations of 259 of
them. In that way, they identified forty seven quasars: thirty eight
behind the LMC and nine behind the SMC. Three of their SMC objects
were on the list of Eyer's candidates: two quasars and a Be star.

In this paper, we present the identifications of the new quasars
behind the SMC, and we discuss the prospects for application of Eyer's
method for quasar searches behind other dense stellar fields.


\section{Observations and identifications}

The optical spectra were obtained on 2002 September 16-18 with the
Magellan Baade 6-5 meter telescope. We used the LDSS-2 imaging
spectrograph, with a 2048$\times$2048 SITe\#1 CCD with a scale of
0.38~arcsec/pixel, a gain of 1~$e^-/$ADU, and a readout noise of
$7e^-$. The slit width was 1.03~arcsec and the grism setting was
300~l/mm, yielding a nominal resolution of 13.3~\AA. Exposure times
ranged from 120 to 600 seconds. All observations were carried out with
the slit oriented in the east-west direction. Additionally, we
observed two spectrophotometric standards, LTT 1788 and LTT 7379
(Hamuy et al.\ 1992). Following each observation, a He-Ne arc lamp
spectrum was acquired for wavelength calibration purposes. Spectra
were reduced in the standard way using IRAF.

\begin{figure}
\epsscale{0.81}
\plotone{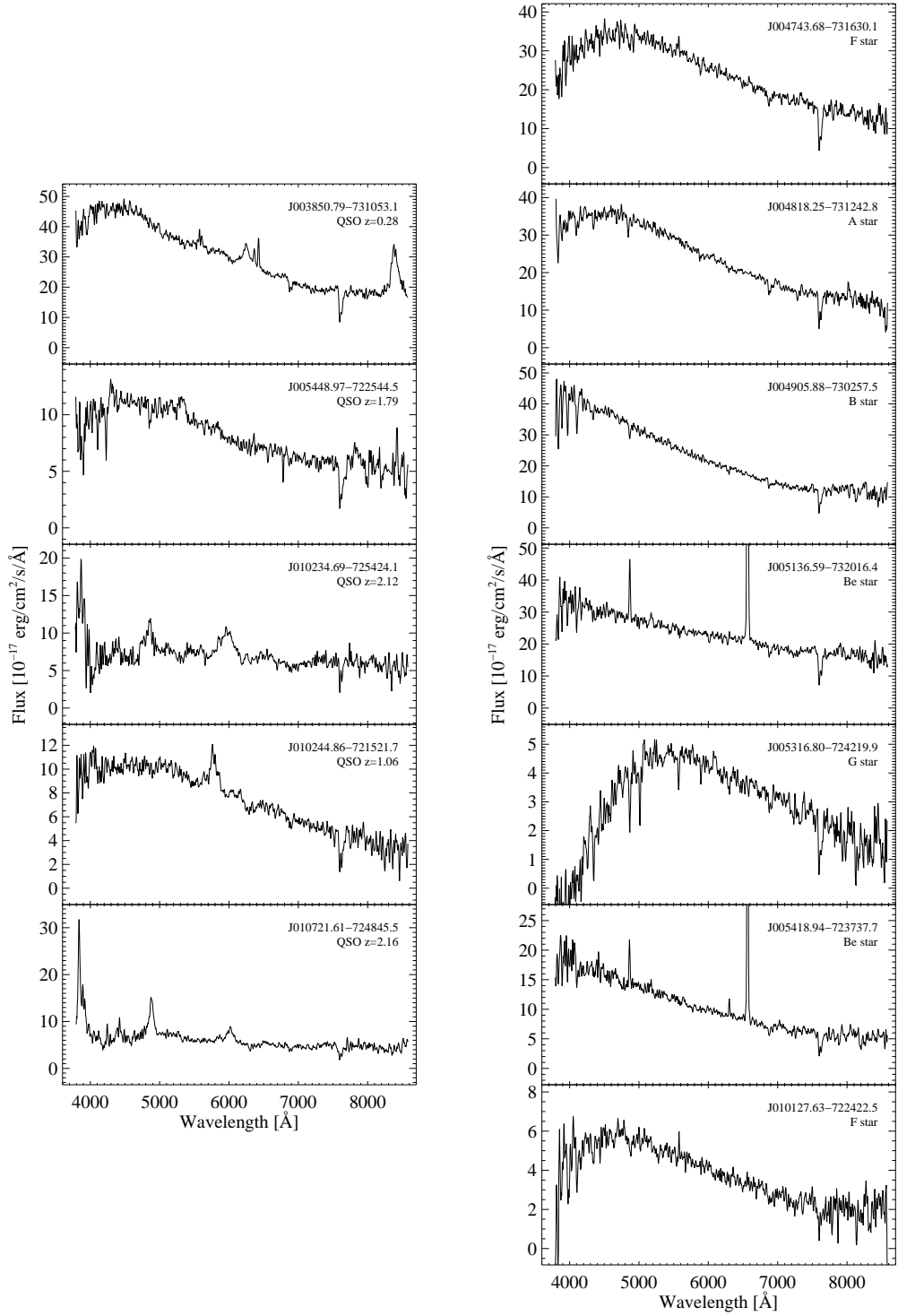}
\caption{Magellan spectra of twelve brightest ``Quasar Candidates''
from Eyer (2002). Left panel shows identified quasars, right panel
shows objects that turned out to be stars.\label{fig:spectra}}
\end{figure}

Figure~\ref{fig:spectra} shows the spectra of twelve of Eyer's QC
objects. There are five confirmed quasars among them, and we show
their spectra on the left panel, while on the right panel one can find
spectra of objects that turned out to be stars. We summarize the
object properties in Table~\ref{tab:identifications}. We also observed
all four of Eyer's ``Unclassified'' objects, which all turned out to
be stars; for completeness, we include this information in
Table~\ref{tab:identifications}.


\begin{figure}
\epsscale{0.81}
\plotone{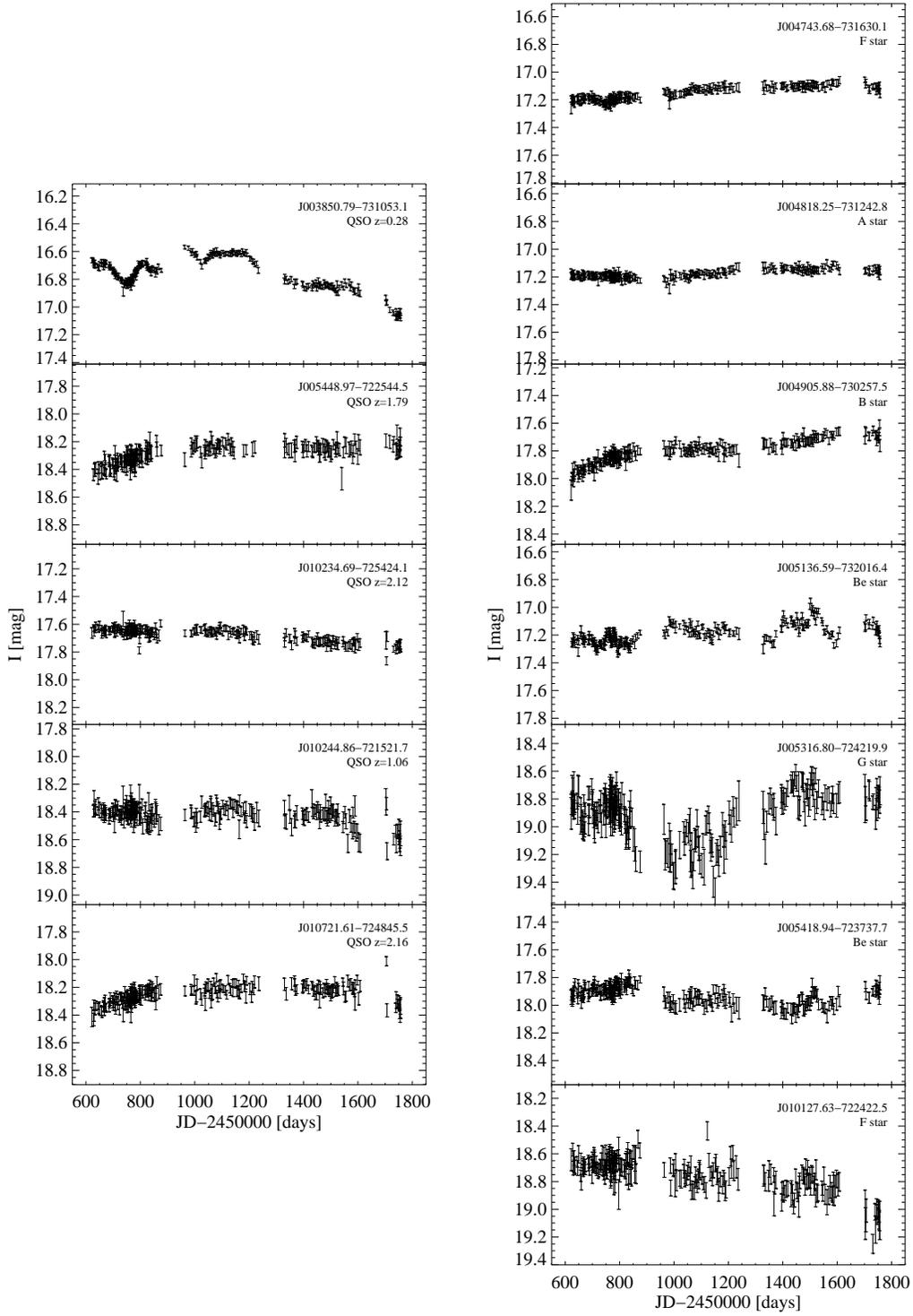}
\caption{OGLE (\.Zebru\'n et al.\ 2001) light curves for objects from
Figure~\ref{fig:spectra}, arranged in a similar way. JD 2,450,000
corresponds to UT 1995 October 9.\label{fig:lightcurves}}
\end{figure}


\begin{deluxetable}{cccccc}
\tablecolumns{6}
\tabletypesize{\scriptsize}
\tablenum{1}
\tablewidth{0pt}
\tablecaption{Object identifications.\label{tab:identifications}}
\tablehead{%
\colhead{OGLE ID\tablenotemark{a}} &
\colhead{$V$\tablenotemark{b}} &
\colhead{$B$\tablenotemark{b}} &
\colhead{$I$\tablenotemark{b}} &
\colhead{Eyer ID} &
\colhead{Identification} \\
\colhead{} &
\colhead{[mag]} &
\colhead{[mag]} &
\colhead{[mag]} &
\colhead{} &
\colhead{}
}
\startdata
\cutinhead{Quasar Candidates} 
003850.79--731053.1 & 17.683 & 17.870 & 16.822 & S1  &  QSO, $\zem=0.28$ \\
004743.68--731630.1 & 17.532 & 17.666 & 17.211 & S3  &  F star           \\
004818.25--731242.8 & 17.446 & 17.493 & 17.193 & S4  &  A star           \\
004905.87--730257.5 & 17.973 & 18.002 & 17.833 & S5  &  B star           \\
005136.59--732016.5 & 18.036 & 18.349 & 17.241 & S6  &  Be star          \\
005316.80--724219.9 & 19.263 & 19.432 & 18.860 & S7  &  G star           \\
005418.96--723737.7\tablenotemark{c}
& 18.137 & 18.194 & 17.893 & S8  &  Be star          \\
005448.97--722544.6 & 19.001 & 19.180 & 18.263 & S9  &  QSO, $\zem=1.79$ \\
010127.64--722422.6 & 19.058 & 19.191 & 18.673 & S13 &  F star           \\
010234.69--725424.1\tablenotemark{c,d}
& 18.346 & 18.689 & 17.640 & S12 &  QSO, $\zem=2.12$ \\
010244.89--721521.7 & 18.892 & 19.385 & 18.412 & S11 &  QSO, $\zem=1.06$ \\
010721.61--724845.5\tablenotemark{c,e}
 & 18.970 & 19.218 & 18.279 & S15 &  QSO, $\zem=2.16$ \\
\cutinhead{Unclassified}
004504.34--724449.9 & 17.906 & 18.245 & 17.356 & S22 &  F star           \\
004702.90--730800.7 & 18.036 & 18.506 & 17.346 & S23 &  F star           \\
005039.12--724154.3 & 18.871 & 19.138 & 18.305 & S24 &  F star           \\
005137.19--731429.2 & 17.221 & 17.240 & 16.973 & S25 &  Be star          \\
\enddata
\tablenotetext{a}{OGLE ID contains J2000.0 equatorial coordinates.}
\tablenotetext{b}{Magnitudes from Udalski et al.\ 1998.}
\tablenotetext{c}{Object identified in MACHO database by Geha et al.\
2002.}
\tablenotetext{d}{Near object 171 in Kahabka et al.\ 1999 and object
182 in Haberl et al.\ 2000.}
\tablenotetext{e}{Near object 203 in Kahabka et al.\ 1999 and object
338 in Haberl et al.\ 2000.}
\end{deluxetable}


The spectra of all five quasars show at least two emission lines,
enabling unambiguous determination of emission redshifts. As mentioned
earlier, two of those quasars (QSO J010234.69--725424.1 and QSO
J010721.61--724845.5) were independently identified by Geha et al.\
(2002).


\section{Discussion}

We have identified five variability-selected quasars among twelve
brightest Quasar Candidates identified by Eyer (2002) based on the
characteristics of their light curves in the OGLE database. The method
is very efficient. As expected, the contaminants in the list of
candidates were predominantly early type stars in the
SMC. Qualitatively, we do not see any obvious trends or differences
between the light curves of quasars and stars. In
Figure~\ref{fig:lightcurves} we show the OGLE light curves of the QC
objects, arranged similarly to Fig.~\ref{fig:spectra}.


\begin{figure}
\plotone{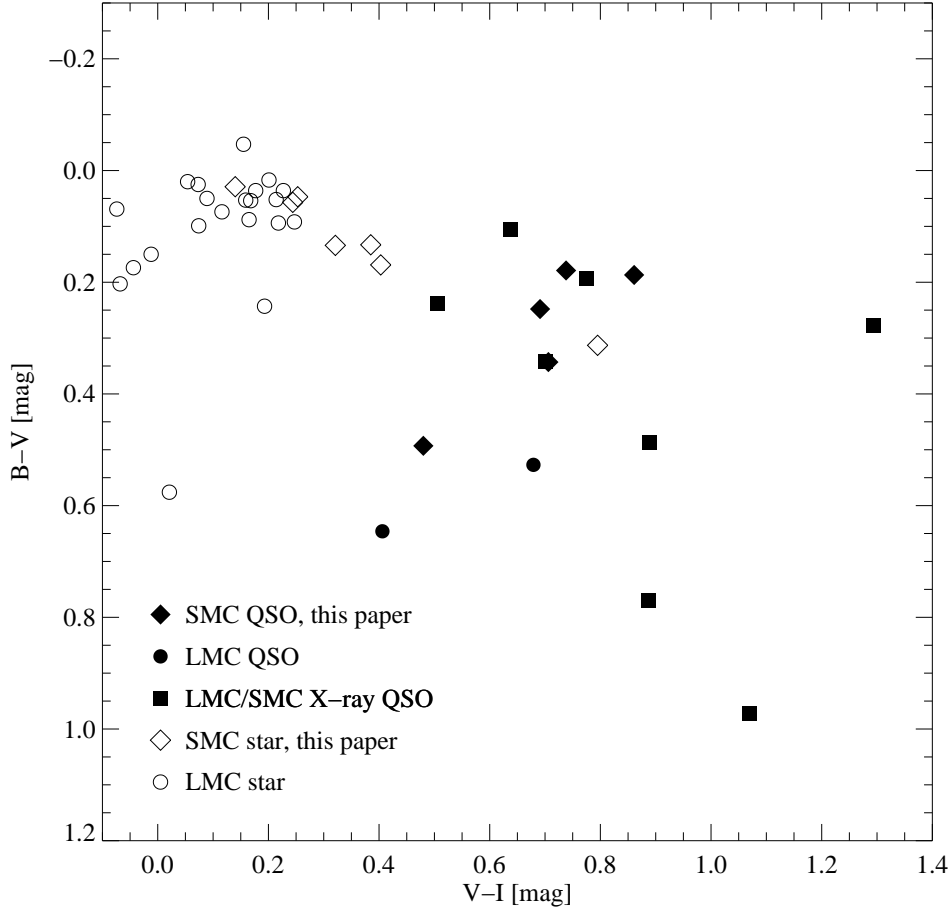}
\caption{The $V-I$ versus $B-V$ color-color diagram for Eyer objects
for which identifications are available. Filled and open diamonds show
quasars and stars, respectively, presented in this paper. The circles
show quasars (filled) and stars (open) that were included in Eyer's
quasar candidate list for the LMC and were identified by Geha et al.\
(2002) and Huchra (2002, private communication). Eyer candidate lists
contain objects for which $V-I<0.9$. We also show (solid squares) the
quasars identified by us via matching OGLE sources with serendipitous
X-ray sources from the {\em Chandra X-ray Observatory\/} (Dobrzycki et
al.\ 2002; 2003, in preparation); Eyer's color criterion was not
applied for those objects.\label{fig:colcol}}
\end{figure}


It appears, however, that the source colors will help in improving the
candidate lists. In Figure~\ref{fig:colcol} we show the color-color
diagram for the twelve sources presented in this paper, plus twenty
four Eyer's QCs from the LMC region which were identified by Geha et
al.\ (2002) and Huchra (2002, private communication). One can clearly
see that variability-selected stars and quasars occupy different
regions in the color-color space. We note here that the Faint Sky
Variability Survey (Groot et al.\ 2002) data seem to show a similar
effect.

In Fig.~\ref{fig:colcol} we also show the known X-ray selected quasars
in both the LMC and the SMC for which OGLE photometry is
available. Those quasars also separate well from the stars. We note
that one of Eyer's selection criteria for quasar candidates was a
color cutoff, $V-I<0.9$, but this criterion was not applied to the
X-ray-selected quasars before they were identified. They are typically
redder than Eyer's objects.

As mentioned earlier, until very recently only a handful of confirmed
quasars were known in the general direction of the SMC (Crampton et
al.\ 1997; Tinney et al.\ 1997). All those quasars are located in
fairly sparse stellar fields away from the center of the SMC, which
limits their applicability to studies of SMC proper motion or
investigations of absorption properties of the SMC. Quasars presented
in the present paper and objects from Geha et al.\ (2002) lie behind
the dense parts of the SMC. Note that those quasars --- extremely
interesting and useful objects in their own right --- are in reality
byproducts of monitoring surveys, unrelated to the original scientific
goals of the surveys. It is an excellent example that such projects
can lead to unexpected, yet very valuable results.

Geha et al.\ (2002) independently identified two of our quasars. We
note that those two objects are the largest redshift quasars among our
five, but this most likely is just a coincidence, not a result of the
difference in the methods applied by Geha et al.\ and Eyer. Geha et
al.'s quasar redshifts span a large redshift range. We note, however,
that two methods, which, after all, are based on a similar concept,
apparently have a rather small intersection in the final candidate
lists: there are only ten objects in common in both the SMC and the
LMC. This fact and the fact that X-ray-selected quasars are redder
than the variability-selected objects indicate that both techniques
are conservative and both will miss some quasars. The two methods
should therefore be considered as complementing one another, rather
than as competing.

We note here that the same two quasars lie relatively close to {\em
ROSAT\/} X-ray sources listed in Kahabka et al.\ (1999) and Haberl et
al.\ (2000), although neither one of the X-ray sources was classified
as a probable QSO.

One of the quasars identified in this paper, QSO J003850.79--731053.1
($\zem=0.28, V=17.7$) is a very promising candidate for studies of
absorption in the SMC. Its brightness should enable good spectroscopy
with the Cosmic Origin Spectrograph aboard the Hubble Space
Telescope. The other four quasars and quasars from Geha et al.\ (2002)
are very well positioned to be reference points for SMC proper motion
studies. We add here that we also identified six other X-ray selected
quasars behind the dense parts of the SMC; we will present them in the
forthcoming paper (Dobrzycki et al.\ 2003, in preparation).

We were somewhat surprised to find a G-type star among Eyer's QC
objects, especially since he did utilize color selection in
constructing the list of candidates. However, it was also the faintest
of the observed objects, for which the photometry is likely rather
uncertain.

Excellent efficiency of the variability-based method bodes well for
searches of quasars behind other dense stellar fields for which
monitoring photometry databases are available. A first successful
search has already been published by Geha et al.\ (2002), who found 38
QSOs behind the LMC in the MACHO database.

The paper by Eyer (2002) contains a list of 118 QSO candidates behind
the LMC. At present, identifications are known for $\sim$25 of them
(Geha et al.\ 2002; Huchra 2002, private communication). As noted
earlier, the color-color diagram for identified Eyer's QSO candidates
indicates that quasars are typically redder in $V-I$ than QCs that
turned out to be stars (Fig.~\ref{fig:colcol}). Remaining objects from
Eyer's LMC candidate list split roughly evenly between the quasar and
stellar regions in the color-color diagram, indicating that the
spectroscopic followup should yield a large number of quasars.

Geha et al.\ (2002) noted that their quasars tend to be bluer when
they brighten. OGLE time coverage in $B$ and $V$ filters is much
sparser than in $I$, which, combined with relatively small size of our
sample, does not allow us to make quantitative statements of that
nature. We do note, however, that this effect is present in the
brightest of our quasars, QSO J003850.79--731053.1: the well defined
features in the light curve of this object (top left panel on
Fig.~\ref{fig:lightcurves}) have corresponding changes in the $V-I$
color consistent with the effect seen in MACHO quasars.

Another extremely interesting region to search for quasars where the
method could be applied is the Galactic Bulge. To our knowledge, no
quasars have been identified so far in the vicinity of the Galactic
center. There are, however, several regions where interstellar
extinction is quite low ($A_V<3$) and where one could, in principle,
see quasars. The best known such area is Baade's Window (e.g.\ Stanek
1996), but there are several other.

Identifying quasars behind the Galactic Bulge would be very
valuable. Recently, Sumi, Eyer, \& Wo\'zniak (2002) have shown that
there is a statistically significant difference between proper motions
of faint versus bright red clump stars in one of the OGLE bulge
fields. Finding fixed reference points for this study, which quasars
could provide, would be an extremely interesting result.

On one hand, the search toward the Galactic bulge will be made easier
by the fact that there will be few early type stars, which in the
Magellanic Clouds are the primary contaminants in the candidate
lists. On the other hand, the quasars will be considerably reddened
even in the low extinction windows, making them less conspicuous as
blue objects. Also, the surface density of objects that need to be
analyzed will be very large, and as a result there will likely be a
sizeable number of artifacts, etc., which will contaminate the
candidate lists.


\acknowledgments

We would like to thank L. Eyer, B. Paczy\'nski and J. Stocke for
helpful discussions, J. Huchra for observing the LMC candidates, and
the referee, M. Geha, for helpful comments and, especially, for her
suggestion to explore the information contained in the color-color
diagram. AD acknowledges support from NASA Contract No.\ NAS8-39073
(CXC). LMM was supported by the Hubble Fellowship grant HF-01153.01-A
from the Space Telescope Science Institute, which is operated by the
Association of Universities for Research in Astronomy, Inc., under
NASA contract NAS5-26555.



\end{document}